\documentclass[%
 reprint,
 amsmath,amssymb,
 aps,
]{revtex4-1}

\usepackage{graphicx}
\usepackage{grffile}
\usepackage{subfigure}
\usepackage{dcolumn}
\usepackage{bm}
\usepackage{hyperref}

\usepackage{amsmath}
\usepackage{amssymb}

\usepackage[section]{placeins}

\begin{document}

\preprint{APS/123-QED}

\title{Adsorption transition of a grafted ferromagnetic filament controlled by external magnetic fields}

\author{Pedro A. S\'anchez}
\email{r.p.sanchez@urfu.ru}
\affiliation{%
 Institute of Natural Sciences and Mathematics, Ural Federal University, Ekaterinburg, Russian Federation
}%
\altaffiliation[Also at ]{Wolfgang Pauli Institute (P.A.S.) and Computational and Soft Matter Physics (S.S.K.), University of Vienna, Vienna, Austria}
\author{Ekaterina V. Novak}
\affiliation{%
 Institute of Natural Sciences and Mathematics, Ural Federal University, Ekaterinburg, Russian Federation
}%

\author{Elena S. Pyanzina}
\affiliation{%
 Institute of Natural Sciences and Mathematics, Ural Federal University, Ekaterinburg, Russian Federation
}%

\author{Sofia S. Kantorovich}
\affiliation{%
 Institute of Natural Sciences and Mathematics, Ural Federal University, Ekaterinburg, Russian Federation
}%

\author{Joan J. Cerd\`a}%
\affiliation{%
Dpt. de F\'isica UIB i Institut d'Aplicacions Computacionals de Codi Comunitari (IAC3), Universitat de les Illes Balears, Palma de Mallorca, Spain 
}%


\author{Tom\'as Sintes}
\affiliation{%
Instituto de F\'isica Interdisciplinar y Sistemas Complejos, IFISC (UIB-CSIC), Universitat de les Illes Balears, E-07122 Palma de Mallorca, Spain 
}



\date{\today}

\begin{abstract}
Extensive Langevin dynamics simulations are used to characterize the adsorption transition of a flexible magnetic filament grafted onto an attractive planar surface. Our results identify different structural transitions at different ratios of the thermal energy to the surface attraction strength: filament straightening, adsorption and the magnetic flux closure. The adsorption temperature of a magnetic filament is found to be higher in comparison to an equivalent nonmagnetic chain. The adsorption has been also investigated under the application of a static homogeneous external magnetic field. We found that the strength and the orientation of the field can be used to control the adsorption process, providing a precise switching mechanism. Interestingly, we have observed that the characteristic field strength and tilt angle at the adsorption point are related by a simple power law.

\end{abstract}

\pacs{Valid PACS appear here}
\maketitle


\section{\label{sec:intro}Introduction}
The combination of polymers and micro- or nanoparticles is one of the most successful available approaches for the design of novel materials with highly tunable properties \cite{2006-balazs, 2019-alexandrov-ptrsa}. One of the main and simpler examples is the filling of polymer matrices with magnetic particles to create magnetoresponsive gels and elastomers whose mechanical properties can be changed on the fly by means of external fields\cite{2007-filipcsei, 2016-odenbach-aam, 2017-lopez-lopez-pre, 2017-lopez-lopez-pa, 2018-shamonin}. At the nanoscale, polymer coatings are broadly used to stabilize magnetic nanoparticles in suspension as an essential ingredient for the synthesis of ferrofluids and the fine tuning of their static and dynamic properties\cite{2006-odenbach, 2009-vekas, 2018-ivanov-pre, 2019-lebedev-pre}.

A more sophisticated approach for the synthesis of hybrid polymer magnetic materials is the polymer crosslinking of assembled magnetic particles in order to stabilize specific structures. The simplest case is the stabilization of the linear chains that magnetic micro- and nanoparticles tend to form under uniform static external fields. Such linear micro- and nanostructures, often addressed as magnetic filaments (MF), can be used in numerous applications that take advantage of their high magnetic response and shape anisotropy \cite{2005-cebers, 2011-wang, 2016-cebers-afm}. For instance, MFs have been used for the design of magnetically actuated artificial propellers and swimmers \cite{2005-dreyfus, 2009-belovs-pre, 2010-benkoski}, micromechanical sensors \cite{2003-goubault}, microchannel actuators and mixers \cite{2009-fahrni, 2020-zaben-sm} or magnetic resonance contrast agents \cite{2008-corr}.

A large part of the aforementioned applications of MFs involve two important aspects: a certain degree of flexibility of the chain backbone and its sensible interaction with rigid surfaces, being both particularly important for micro- and nanofluidic applications. Regarding the first aspect, current synthesis techniques allow polymer crosslinking of paramagnetic microparticles with a high control of the degree of flexibility of the resulting filament \cite{2014-byrom}. Due to their smaller size, control on the crosslinking of monodomain ferromagnetic nanoparticles is much more difficult and most attempts to date have achieved rather rigid structures only \cite{2007-benkoski-jacs, 2009-zhou, 2014-hill}. However, it has been shown already that it is possible to create flexible noncrosslinked chains of polymer coated ferromagnetic nanoparticles \cite{2014-townsend} and cutting-edge synthesis techniques, such as polymer templating \cite{2012-sarkar} or DNA directed self-assembly \cite{2015-tian-nn, 2016-tian-nm}, are paving the way to the creation of highly flexible nanofilaments of monodomain superparamagetic and ferromagnetic nanoparticles. Interactions of MFs with surfaces are also widespread among their technological applications. In many cases, the filaments are end-grafted to the surface of a larger particle (for instance, in artificial swimmers) or wall (pumpers, mixers). In addition, the structural similarity of MFs with molecular polymers has inspired their use in dense polymer brush-like arrangements in order to create magnetoresponsive coatings \cite{2008-choi, 2010-ye, 2015-sanchez-mm1, 2016-sanchez-fd, 2017-pyanzina-sm, 2019-cerda-sm}. To this regard, permanently stabilized and grafted nanoscale flexible filaments can broaden the already promising potential for applications of brush-like systems of simple magnetically assembled chains, hybrid polymer microfibers or rigid magnetic micropillars\cite{2010-vilfan, 2014-tokarev, 2015-sun-fms, 2016-orlandi-pre, 2018-hanasoge-mn}.

Following many of the applications mentioned above, most theoretical studies on surface grafted MFs to date have been focused on their magnetoelastic response and hydrodynamic interactions with the background fluid, whereas the grafting surface only played the role of an inert geometric constraint \cite{2003-goubault, 2009-fahrni, 2009-gauger, 2013-chen, 2011-babataheri, 2017-vazquez-montejo-prm, 2017-dempster-pre}. However, one can think in very interesting applications involving noninert surfaces. For instance, attractive surfaces in the walls of a microfluidic channel can adsorb substances carried by the flowing liquid, acting as a filter. The presence of grafted MFs also experiencing the attraction to the surface would provide a switching mechanism on the adsorbing properties of the walls: in absence of external fields the filaments would remain adsorbed, becoming a steric barrier for the adsorption of free flowing components; application of adequate external fields that could force the desorption of the MFs under given flow conditions would activate the adsorption of the former. In order to design a system as such, first it is essential to understand the adsorption process of a grafted MF on an attractive surface and how this can be controlled by means of external fields. This is the main goal of this work.

Several years ago we presented the first theoretical study on the adsorption transition of a free semiflexible MF on a flat surface in absence of external fields \cite{2011-sanchez-sm}. Even closely related to the topic studied here, to our best knowledge the field-induced adsorption/desorption of MFs has never been addressed before.

Here we employ computer simulations with a mesoscale model to study the equilibrium behavior of a flexible filament made of ferromagnetic particles grafted to an attractive flat surface. By means of molecular dynamics in the canonical ensemble, we first study the adsorption transition that takes place as the ratio of the thermal fluctuations to the strength of the surface attraction decreases, discussing how it compares to the adsorption of an equivalent nonmagnetic filament. Second, we study how such transition is affected by the application of homogeneous magnetic fields of different strengths and orientations with respect to the surface plane. We show that both, the strength and the orientation of the field can be used to drive the adsorbed or desorbed state of the filament, providing a precise switching mechanism.

\section{\label{sec:model}Simulation model and methods}
\begin{figure}[h]
 \centering
 \includegraphics[width=5.5cm]{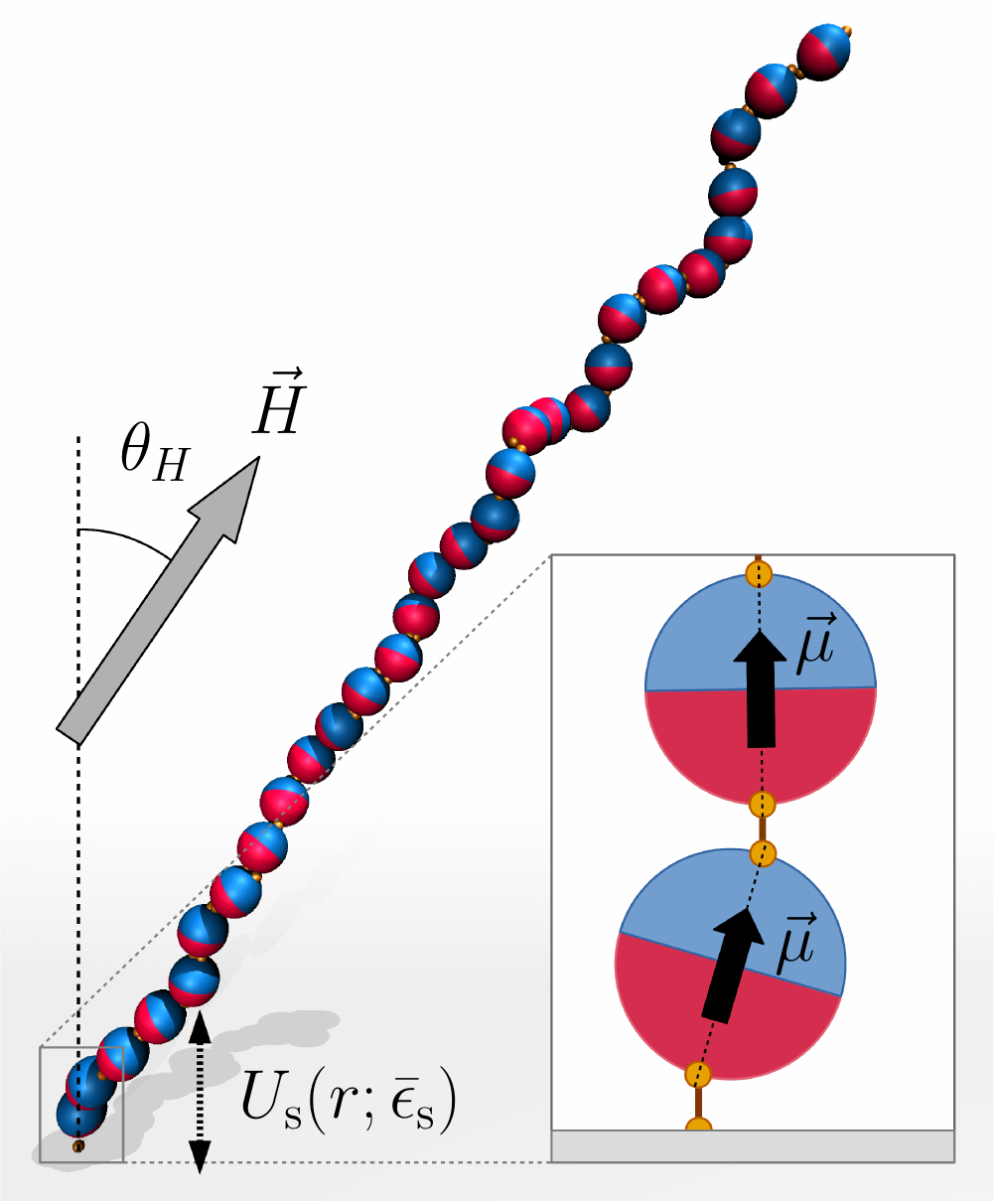}
 \caption{Scheme of the grafted magnetic filament model, showing a configuration equilibrated under a strong applied field, $\vec H$, tilted an angle $\theta_H$ with respect to the normal of the grafting surface. Ferromagnetic particles of the filament are represented as two-color spheres, with colors indicating the orientation of their central dipole moment. See the main text for further details.}
 \label{fig:model}
\end{figure}
Due to the characteristic lengths and time scales involved in hybrid materials that combine polymers and micro- or nanoparticles, simulation models for such systems have to rely on coarse-grained approximations. 
Numerous computer simulations of magnetic gels, elastomers and filaments are based on bead-spring representations with different levels of detail \cite{2011-wood-pre, 2013-annunziata-jcp, 2014-tarama-pre, 2015-weeber-jmmm, 2016-pessot-jcp, 2018-weeber-jpcm}. Regarding particle-based simulation of MFs, since the main role of the polymer components is to provide the permanent linking of the magnetic particles in the chain, the most convenient approach is to represent them implicit by means of simple elastic bonding potentials, whereas the particles are usually simulated as beads with point magnetic dipoles \cite{2011-sanchez-sm, 2015-pshenichkov-pre, 2016-wei-lm, 2018-kuznetsov-jmmm, 2020-mostarac-jml}. Our mesoscale model of the grafted ferromagnetic filament is based on such approach and is very similar to those used in our previous studies on these systems \cite{2011-sanchez-sm, 2015-sanchez-mm1, 2016-sanchez-fd, 2017-pyanzina-sm, 2019-cerda-sm}.

Figure~\ref{fig:model} shows a sketch of our model system and its interactions. We consider only the case of a filament formed by $N_p=30$ identical ferromagnetic particles. The latter are modeled as spherical beads of characteristic diameter $\sigma$, carrying a point magnetic dipole $\vec \mu$ at their centers. Note that this is an accurate description for spherical monodomain ferromagnetic particles. In addition, we adopt the limit of infinite magnetic anisotropy for these particles, so that their dipole moments have not only a constant modulus but also a fixed orientation with respect to the particle's body frame. 

The steric interactions between these particles, as well as their attractive interaction with the grafting surface, are represented by means of truncated Lennard-Jones (LJ) type potentials, shifted to make them smoothly vanish at the truncation point, $r_{\mathrm{cut}}$:
\begin{multline}
U_{\mathrm{{t-s}}}(r; \epsilon, \sigma, r_{\mathrm{cut}})=\\
=\left\{ \begin{array}{ll}
U_{\mathrm{{LJ}}}(r; \epsilon, \sigma)-U_{\mathrm{{LJ}}}(r_{\mathrm{cut}}; \epsilon, \sigma), & r<r_{\mathrm{{cut}}}\\
0, & r\geq r_{\mathrm{{cut}}}
\end{array}\right. .
\label{eq:LJts}
\end{multline}
where $\epsilon$ is the energy scale of the interaction and $r$ is the characteristic separation distance (center-to-center for particle pairs or center-to-surface for particle-surface interactions). In order to model a purely repulsive steric interaction between the particles as the one produced by a soft polymer coating, we take the conventional Lennard-Jones potential,
\begin{multline}
U_{\mathrm{LJ}}(r; \epsilon_{\mathrm{c}}, \sigma)= \\
= U^{12-6}_{\mathrm{LJ}}(r; \epsilon_{\mathrm{c}}, \sigma) = 4 \epsilon_{\mathrm{c}} \left [ \left ( \frac{\sigma}{ r}\right )^{12} - \left ( \frac{\sigma} {r}\right )^{6} \right ],
\label{eq:LJ}
\end{multline}
truncated at the position of its minimum, $r_{\mathrm{cut}} = 2^{1/6} \sigma$. The combination (\ref{eq:LJts}) and (\ref{eq:LJ}) corresponds to the soft-core interaction known as Weeks-Chandler-Andersen (WCA) potential \cite{1971-weeks}. For the interaction with the surface we apply expression~(\ref{eq:LJts}) to a 9--3 LJ potential,
\begin{multline}
U_{\mathrm{LJ}}(r; \epsilon_{\mathrm{s}}, \sigma)= \\
= U^{9-3}_{\mathrm{LJ}}(r; \epsilon_{\mathrm{s}}, \sigma) = \frac{3\sqrt{3}}{2} \epsilon_{\mathrm{s}} \left [ \left ( \frac{\sigma}{ r}\right )^{9} - \left ( \frac{\sigma} {r}\right )^{3} \right ],
\label{eq:LJ93}
\end{multline}
which is the result of integrating the conventional potential~(\ref{eq:LJ}) over an infinite plane. In this case we make the interaction attractive by taking $r_{\mathrm{cut}} = 3.5 \sigma$. Despite the truncation and shifting of attractive Lennard-Jones type potentials introduces a discontinuity in their derivatives at the truncation point, the use of such large cutoff, correponding to the maximum value in the range most frequently used in simulations, provides a discontinuity small enough to ensure that its effects will be negligible in front of the thermal fluctuations, at least for the range of temperatures of interest sampled here. Note that the resulting potential, which we label as $U_{\mathrm{s}}$, has a well whose minimum is located at $r_{\mathrm{min}}=3^{1/6} \sigma$ and, after applying the shift corresponding to the selected truncation, its depth is $\bar \epsilon_{\mathrm{s}} = -\epsilon_{\mathrm{s}} \left [ 1 - 3\sqrt{3}(3.5^{-9} - 3.5^{-3} )/2\right ] \approx -1.06 \epsilon_{\mathrm{s}}$. In the following we will discuss the strength of the attraction to the surface in terms of $\bar \epsilon_{\mathrm{s}}$.

We assume the chain structure of our filament to be stabilized by long polymer crosslinks attached to a very narrow region of the surface of the linked particles. Under these conditions the filament backbone is flexible and the crosslinks can be modeled as a simple finitely extensible nonlinear elastic (FENE) potential. This bonding potential is defined as \cite{1986-grest}
\begin{equation}
 U_{\mathrm{FENE}}(r; K, r_{\mathrm{max}}) = -\frac{1}{2} K r_{\mathrm{max}}^2  \ln \left [ 1 - \left ( \frac{r}{r_{\mathrm{max}}} \right ) ^2\right ] ,
\end{equation}
where $K$ defines the elastic strength of the bond and $r_{\mathrm{max}}$ its maximum extension. As shown in the sketch of Figure~\ref{fig:model}, the FENE springs are attached to points of the particle's surfaces located at the projections of the head and the tail of their central dipoles. This corresponds to the crosslinking of a chain of particles assembled into a head-to-tail configuration by the presence of a homogeneous external field. Therefore, the orientation of the dipole moments of the particles is coupled to the filament backbone due to the crosslinks.

Each pair of magnetic dipoles $\vec{\mu}_{i}$ and $\vec{\mu}_{j}$ experiences the conventional long-range dipole-dipole pair interaction
\begin{equation}
U_{\mathrm{dip}}(\vec r_{ij};\ \vec \mu_i,\ \vec \mu_j)=\frac{\vec{\mu}_{i}\cdot\vec{\mu}_{j}}{r^{3}}-\frac{3\left[\vec{\mu}_{i}\cdot\vec{r}_{ij}\right]\left[\vec{\mu}_{j}\cdot\vec{r}_{ij}\right]}{r^{5}},
\label{eq:dipdip}
\end{equation}
where $r = \| \vec r_{ij} \|$ is the displacement vector between the the centers of the corresponding particles. Finally, each dipole moment also experiences a Zeeman interaction with applied external fields. In general, the interaction of a point magnetic dipole with a net magnetic flux density $\vec B$ at the position of the former is $U_{\mathrm{Z}}(\vec \mu, \vec B)=-\vec \mu \cdot \vec B$. Here, however, we only need to consider the contribution to $\vec B$ of an applied external field of strength $\vec H$. The approximations assumed in the modeling of the magnetic properties of our particles (point dipoles of fixed modulus, independent of the applied field, and infinite magnetic anisotropy) allows us to simply write
\begin{equation}
 U_{\mathrm{Z}}(\vec \mu, \vec H)=-\vec \mu \cdot \vec H,
 \label{eq:zeeman}
\end{equation}
provided a convenient unit rescaling is used for these parameters. In this way, in Eqs.~(\ref{eq:dipdip}) and (\ref{eq:zeeman}) we expressed the magnetic interactions in our system in terms of an external control parameter, the applied field intensity $\vec H$, and an extensive effective parameter, the dipole moment $\vec \mu$, which incorporates the specific properties of the material forming the particles.

As is usual in simulations with mesoscale models, we use a set of reduced, \textit{i.e.}, dimensionless units. By choosing scales that keep the numerical values not too far from unity the stability of the calculations is enhanced and, importantly, in this way the same model may represent very different systems, as long as the ratios between the distinct interaction strengths remain the same. Here we define lengths and masses in units of the diameter and mass of the beads, so we take $\sigma=1$ and $m=1$. Energy scale, $\epsilon^*$, is given by the strength of the thermal fluctuations at room temperature, $T^*$, so that $\epsilon^*=kT^*$, being $k$ the Boltzmann constant. In order to simplify the notation, henceforth we will use the reduced temperature, $T$, to represent the strength of the thermal fluctuations. Therefore, $T=1$ under room conditions. Time scale is related to the parameters above as $\tau^* = \sigma^* (m^* / \epsilon^*)^{1/2}$, however, since here we are only interested in equilibrium properties, this scale is not relevant for the discussion. The strength of the dipole-dipole interaction is defined naturally by the squared dipole moment of the particles, which we set either to $\mu^2=0$ (nonmagnetic particles) or $\mu^2=5$. Note that scales of dipole moment and applied field strengths are defined by $(4\pi\epsilon^* (\sigma^*)^3/\mu_0)^{1/2}$ and $m^* (\tau^*)^{-2}(\mu_0 \sigma^* /4\pi\epsilon^*)^{1/2}$, respectively. The parameters of the bonding interaction are set to $K=30$, $r_{\mathrm{max}}=0.5$, which provide an average center-to-center distance between linked particles of approximately $\sim$1 at $T=1$. We sample strengths of the attraction to the surface within the interval $\bar \epsilon_{\mathrm{s}} \in [0.14,\, 2.89]$. Finally, we analyze the influence of the temperature and external field on the adsorption of the filament on the grafting surface by sampling respective ranges of temperatures and external field strengths $T\in [0.25,\, 5]$ and $H \in [0, 2] $. This choice of parameters could correspond, for instance, to filaments made of magnetite nanoparticles of $\sim$35~nm in diameter coated with a repulsive soft layer of $\sim$6.5~nm and exposed to external fields of up to $\sim$3~kA/m.

The parameters described above have been sampled by performing molecular dynamics (MD) simulations with a Langevin thermostat. The latter treats implicitly the effects of the thermal fluctuations of the background fluid by introducing stochastic forces and friction terms, satisfying the conventional fluctuation-dissipation rules, in the translational and rotational Newtonian equations of motion \cite{1986-grest, 1987-allen}. The latter have been integrated by means of a velocity Verlet scheme. Since here we are interested in equilibrium properties only, hydrodynamic interactions have not been taken into account.

Despite the simplicity of the system studied here, statistical sampling of transitions of polymer-like structures is in general rather demanding. In order to efficiently improve the statistics, we used the replica exchange molecular dynamics (REMD) method \cite{1999-sugita, 2001-mitsutake}. In this approach, $N$ independent simulations of the same system (replicas) are run in parallel, each with one value of a given parameter, $A$, from an ordered set: $A=\{ A_1,\, A_2,\, \dots,\, A_N\}$, where $A_1 < A_2 < \dots < A_N$. After equilibration of each replica, an attempt to exchange the configurations with adjacent values $A_i$, $A_{i+1}$ is performed according to the Boltzmann probability $P(A_i,\, A_{i+1})$, defined as
\begin{multline}
P(A_i,\, A_j) =
\min\left (1,\, \exp \left [\frac{U_{i}(A_{i})}{A_{i}} +  \right. \right.\\ + \frac{U_{i+1}(A_{i+1})}{A_{i+1}}
\left. \left. - \frac{U_i(A_{i+1})}{A_{i+1}} - \frac{U_{i+1}(A_{i})}{A_{i}} \right ] \right )
\label{eq:exchangeprob}
\end{multline}
where $U_a(A_b)$ is the potential energy of the configuration equilibrated under $A=A_a$ but calculated by taking $A=A_b$. This exchange procedure, which is intended to prevent the system to get trapped into local minima, requires the energy histograms of adjacent replicas to have a significant overlap in order to be effective. However, in general one wants to run the least possible number of replicas that span the range of interest of $A$ in order to minimize the computational load. As a reasonable compromise, we chose sets of parameter values that provided overlaps of about 30\% of the histograms area. Even though REMD simulations are mainly used to simulate systems under different temperatures, this technique can be used to study any parameter affecting the internal energy. In reference \cite{2013-sanchez-a} we used REMD to study the influence of the dipole moment on the equilibrium configurations of a filament in bulk. Here, we applied this approach for the sampling of the sets of temperatures and external fields mentioned above. Specifically, we performed REMD simulations separately for different temperatures at zero field and for different fields strengths and orientations at $T=1$.

REMD simulations are naturally well suited for an additional statistical refinement: the weighted histogram analysis method (WHAM) \cite{1989-ferrenberg-prl, 1992-kumar}. This technique combines statistics from simulations at different values of the parameters by weigthing them according to their thermodynamic probability. Its most widespread application corresponds to the combination of statistics from a set of canonical simulations performed at different inverse temperatures. In our notation this corresponds to $\beta_i=T_i^{-1}=\lbrace \beta_1, \,, \dots,\, \beta_N \rbrace$. After equilibration, each simulation provides a set of $M_i$ measures of the internal energy of the system, $E$, with a correlation length $\tau_i$. The histograms of such measures are estimates of the probability distributions of energy values:
\begin{equation}
    p_i(E) = \frac{h_i(E)}{ \Delta E M_i},
\end{equation}
where $h_i(E)$ is the number of measurements of energies within the interval $[E-\Delta E /2,\, E+\Delta E/2)$. The true canonical distribution is actually
\begin{equation}
    p_i(E) = g(E) e^{-\beta_i E + f_i},
\end{equation}
where $g(E)$ is the density of states, a probability density function that describes the number of configurations with energy $E$ that the system may adopt, and $f_i= - \log Z_{\beta_i}$ is the dimensionless free energy. Therefore, each simulation provides an estimate of the density of states. Individually, such estimates are only accurate within the limited range of energies on which the corresponding histogram has significant values. However, it is possible to make a weighted combination of all the estimates in order to obtain a better approximation:
\begin{equation}
    g(E) = \sum_{i=1}^N w_i(E) p_i(E) e^{-\beta_i E + f_i},
\end{equation}
where the weights $w_i(E)$ should fulfill $\sum_{i=1}^N w_i(E) =1$ for all $E$. Such weights have to be chosen to minimize the uncertainty associated to the histograms. In the simplest scheme, the minimization assumes a Poisson distribution for the histograms and takes into account the correlations of the measures, leading to the pair of expressions
\begin{equation}
 g(E) = \frac{\sum_{i=1}^N l_i h_i(E)}{\sum_{j=1}^N \Delta E M_j l_j e^{-\beta_jE + f_j}}
\end{equation}
and
\begin{equation}
 e^{-f_i} = \sum_E g(E) e^{-\beta_i E},
\end{equation}
where $l_i = (1+2\tau_i)^{-1}$. This set of equations can be solved numerically to determine $g(E)$, for instance by self-consistent iteration. Once the density of states is known, expectation values of any observable of the system, $\langle O \rangle$, at any inverse temperature $\beta^*$ can be calculated as
\begin{equation}
 \langle O \rangle_{\beta^*} = \frac{\sum_E g(E) O(E) e^{-\beta^* E}}{\sum_E g(E) e^{-\beta^* E}}.
\end{equation}
Note that this expression will provide good estimations for any arbitrary $\beta^*$ within the range of sampled values, \textit{i.e.}, its estimations are not limited to the discrete set of temperatures used in the simulations. This method will be applied in next Section to obtain finely resolved curves of the adsorption energy and the structural parameters of the filament as a function of the temperature.

The simulation protocol consisted of different MD steps. First, random initial configurations of the grafted chain were prepared for each replica by performing $5\cdot10^6$ integrations of damped dynamics in  absence of magnetic interactions and at temperature $T=2$. The damping, achieved by setting the translational and rotational friction constants, $\Gamma_T$ and $\Gamma_R$, to $\Gamma_T=\Gamma_R=50$, helps to fastly relax artificial initial configurations without the need of a very small time step. The latter was fixed for the whole protocol at $\delta t=10^{-3}$. At this point is important to underline that, since we are only interested in equilibrium properties, the choice of $\Gamma_T$ and $\Gamma_R$ is physically irrelevant and, therefore, it can obey to considerations of simplicity and computational efficiency. Except for the initial damped MD step, we chose to fix $\Gamma_T=\Gamma_R=1$ for the subsequent cycles, as these values are known for providing a fast equilibration in this type of systems. After such initial setup, the final parameters for each replica (temperature and external field) were set, the calculations of the magnetic interactions were enabled and a large set of main equilibration-measures-exchange cycles was started. In each of these cycles, the equilibration consisted in $10^6$ integrations. Measures of the system configurations from each replica were stored during the next $2\cdot10^6$ integrations at intervals of $5\cdot10^5$ integrations. This large amount of integrations between measures ensures small correlations even at low temperatures. Finally, the attempt of configuration exchange for adjacent replicas was carried out. For each simulation set, at least $10^3$ cycles of equilibration-measures-exchange were performed. All the simulations were made with the simulation package {ESPResSo} 3.3.1 \cite{2013-arnold}.

\section{\label{sec:results}Results and discussion}
In the next two Sections we present the simulation results of the equilibrium properties of our model MF grafted to an attractive surface. First, we characterize its adsorption transition on cooling in absence of external field, comparing its behavior to the case of an equivalent nonmagnetic chain. Next, we analyze how the transition of the MF can be favored or hindered by a homogeneous static external field depending on its strength and orientation.

\subsubsection{Adsorption transition at zero field}
Computer simulations allow to easily access every component of the system internal energy. Therefore, we can characterize the adsorption transition by computing the normalized total energy of interaction with the surface,
\begin{equation}
 \bar U_{\mathrm{s}} = \frac{1}{N_p \bar \epsilon_\mathrm{s}} \sum_{i=1}^{N_p} U_{\mathrm{s}}(r_i; \bar \epsilon_{\mathrm{s}}),
\end{equation}
\begin{figure}[t]
 \centering
 \subfigure{\label{fig:ads-M5}\includegraphics[width=8cm]{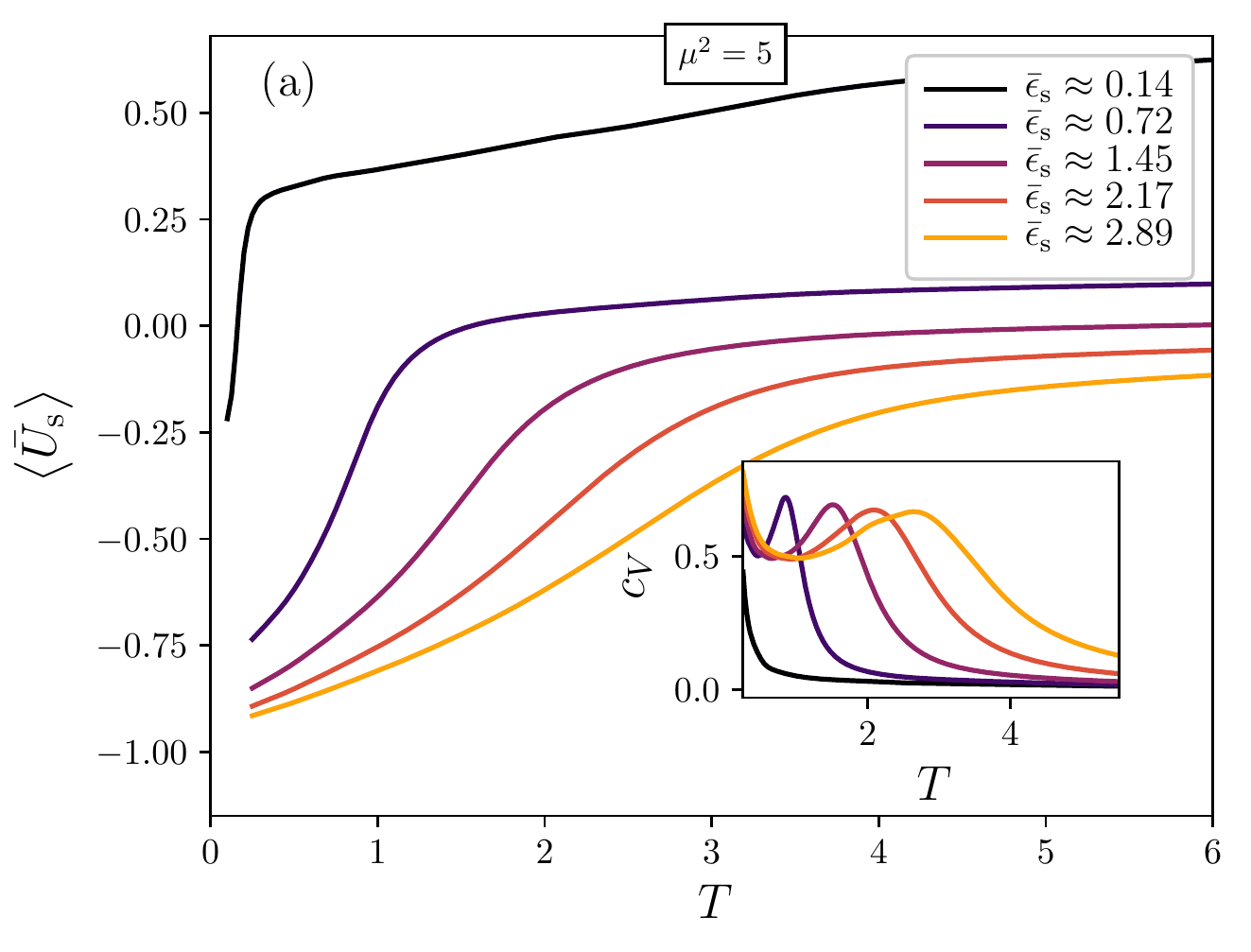}}
 \subfigure{\label{fig:ads-diagram}\includegraphics[width=7.8cm]{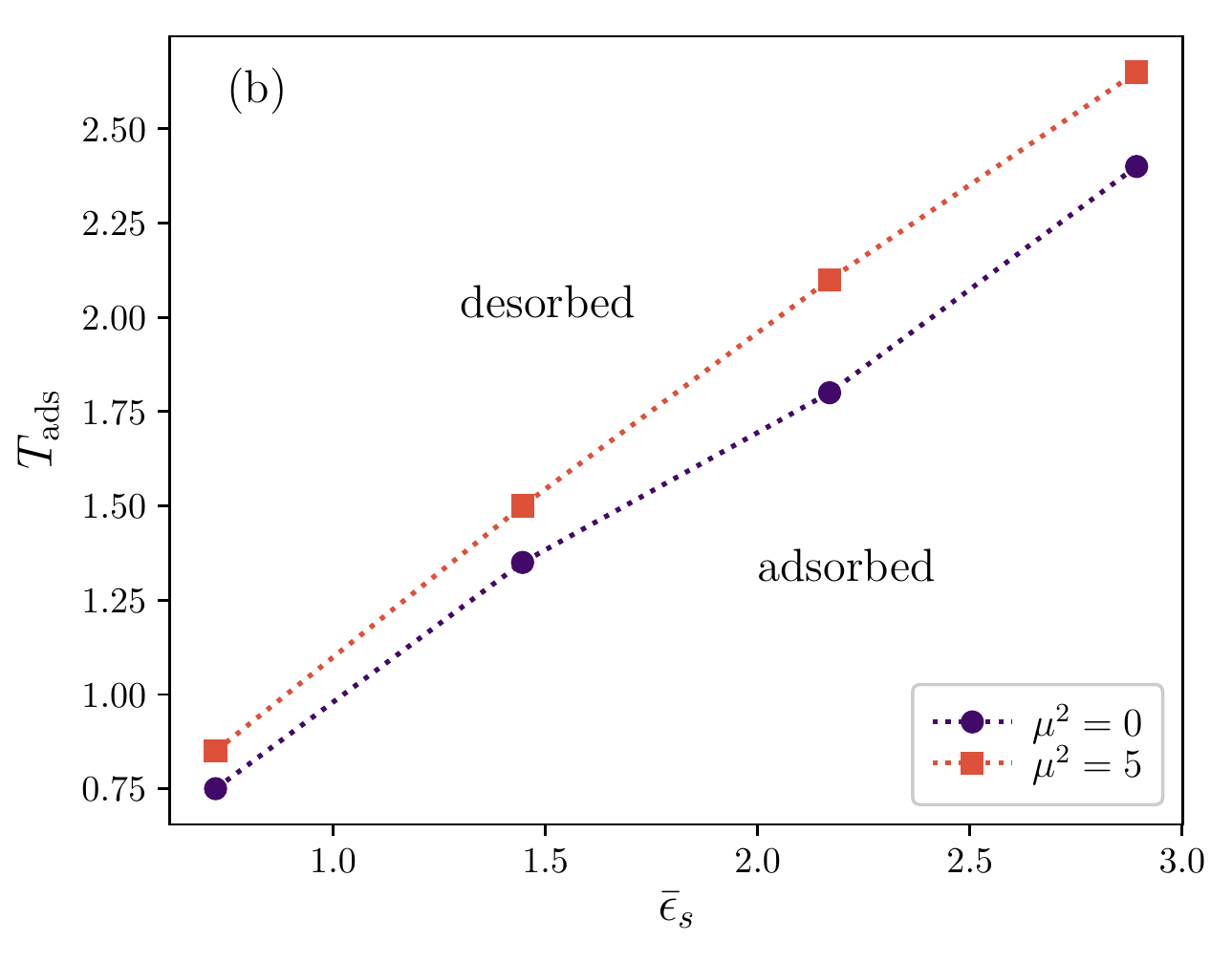}}
 \caption{(a) Surface adsorption transition on cooling for different strengths of surface attraction, as signaled by the surface energy and its corresponding scaled fluctuations (inset), corresponding to magnetic filaments of particles with squared dipole moment $\mu^2=5$. (b) Characteristic temperatures of the adsorption transition for magnetic ($\mu^2=5$) and nonmagnetic ($\mu^2=0$) filaments. Error bars are of the order of the symbol size. Dotted lines are a guide for the eye.}
 \label{fig:adsorption}
\end{figure}
where the sum applies over each particle forming the filament and $U_{\mathrm{s}}(r_i; \bar \epsilon_{\mathrm{s}})$ is given by Eqs.~(\ref{eq:LJts}) and (\ref{eq:LJ93}). Note that the normalization makes this parameter to have a strict lower boundary $\bar U_{\mathrm{s}} \ge -1$. Another useful quantity we can compute is the ratio of the fluctuations of the adsorption energy to the squared thermal energy, normalized by the number of filament beads,
\begin{equation}
 c_V = \frac{\langle \bar U^2_{\mathrm{s}}\rangle - \langle \bar U_{\mathrm{s}}\rangle^2 }{N_p T^2}.
\end{equation}
We label such quantity as $c_V$ due to its analogy with a specific heat. We expect $c_V$ to have a maximum at the characteristic temperature of each transition.

\begin{figure*}[!ht]
 \centering
 \includegraphics[width=18.1cm]{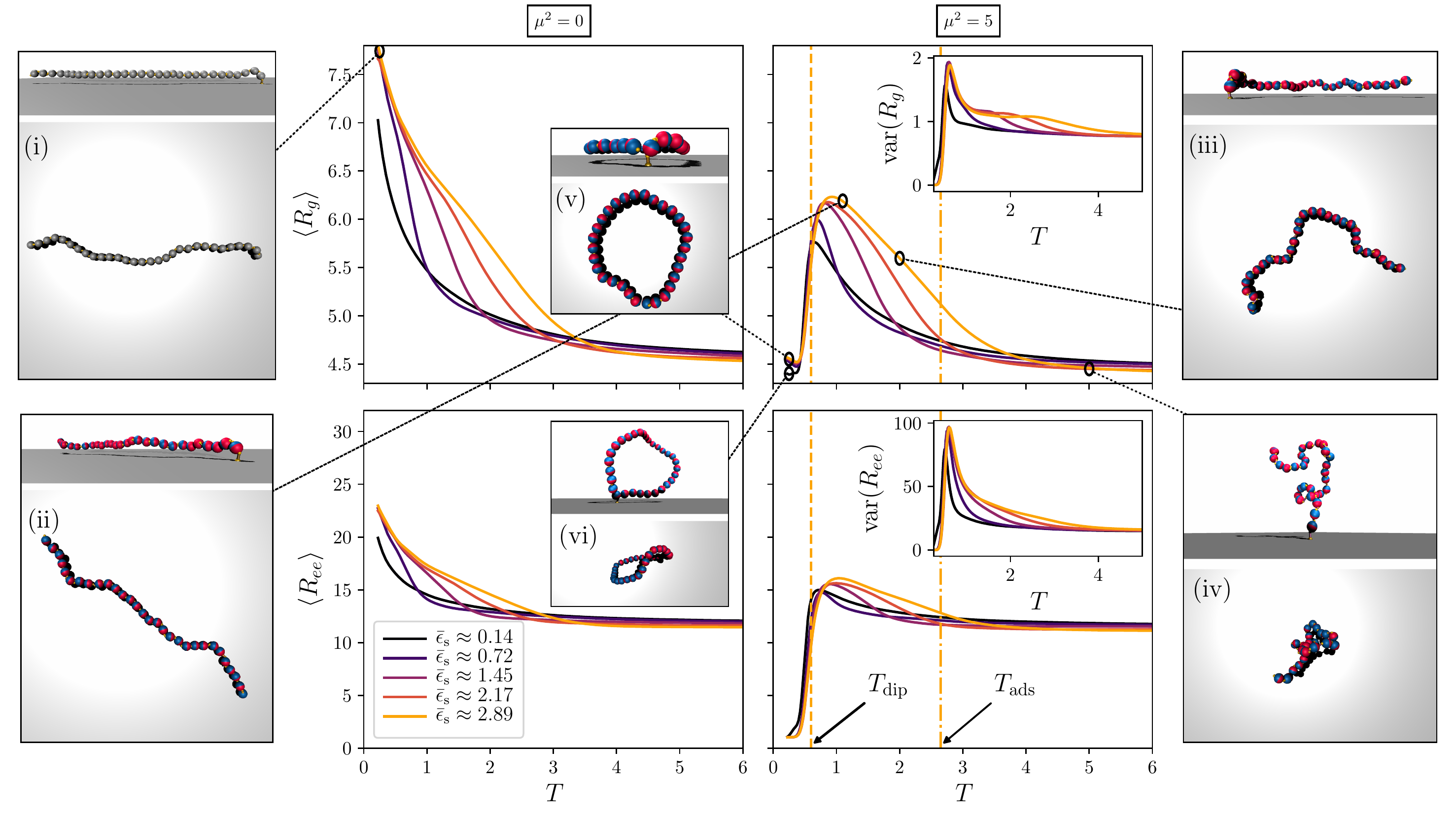}
 \caption{Average radius of gyration ($\langle R_g \rangle$, upper row) and end-to-end distance ($\langle R_{ee} \rangle$, lower row) for nonmagnetic ($\mu^2=0$, left column) and magnetic ($\mu^2=5$, right column) filaments for different strengths of surface attraction. In the right column, insets show the variances of each parameter, dashed and dashed-dotted vertical lines indicate respectively the characteristic temperatures of the adsorption transition, $T_{\mathrm{ads}}$, and the magnetic closure transition, $T_{\mathrm{dip}}$, corresponding to $\bar \epsilon_{\mathrm{s}} \approx 2.89$. A selection of simulation snapshots obtained, except otherwise indicated, at $\bar \epsilon_{\mathrm{s}} \approx 2.89$ are also included (side view in the upper part of each panel, top view in the lower one): (i) adsorbed nonmagnetic filament at $T=0.25$; (ii) adsorbed open MF at $T=1.1$; (iii) adsorbed open MF at $T=2$; (iv) nonadsorbed open MF at $T=5$; (v) adsobed closed MF at $T=0.25$; (vi) nonadsorbed closed MF at $T=0.25$ and $\bar \epsilon_{\mathrm{s}} \approx 0.14$.}
 \label{fig:rads-nofield}
\end{figure*}
Figure~\ref{fig:ads-M5} shows adsorption transition curves of the magnetic chain ($\mu^2=5$) for different strengths of surface attraction, $\bar \epsilon_{\mathrm{s}}$. They correspond to the average surface energy, $\langle \bar U_{\mathrm{s}} \rangle$, measured as a function of the temperature by means of WHAM calculations on the simulation data. We can see that at high temperatures $\langle \bar U_{\mathrm{s}} \rangle$ tends to display a plateau whose value increases with decreasing $\bar \epsilon_{\mathrm{s}}$. Under such conditions the thermal fluctuations are too strong to let the adsorption take place and the most entropically favorable configurations for the filament are those that minimize its contact with the surface. However, the grafted end particle is constrained to remain within its interaction range in any case. This produces the $\bar \epsilon_{\mathrm{s}}$-dependent bias in the surface energy observed for the high temperature limit. With decreasing $T$, the onset of the adsorption transition can be identified by the corresponding drop of $\langle \bar U_{\mathrm{s}} \rangle$, being more abrupt and taking place at lower temperatures as the surface interaction strength weakens. The scaled fluctuations corresponding to each curve, shown in the inset of Figure~\ref{fig:ads-M5}, have a well defined peak in all cases but for the weakest surface attraction. The same exception can be observed in the trend of the main curves within the region of low temperatures: $\langle U_{\mathrm{s}} \rangle$ tends to approach its lower boundary in all cases except for $\bar \epsilon_{\mathrm{s}} \approx 0.14$, which is significantly lower than the lowest sampled strength of thermal fluctuations, $T=0.25$. This simply reflects the fact that the adsorption transition can only take place at $T$ values comparable to $\bar \epsilon_{\mathrm{s}}$, which can be evidenced by obtaining the characteristic adsorption temperature, $T_{\mathrm{ads}}$, corresponding to each $\bar \epsilon_{\mathrm{s}}$ from the positions of the peaks of $c_V$.

The dependence of $T_{\mathrm{ads}}$ on $\bar \epsilon_{\mathrm{s}}$ for both, magnetic and nonmagnetic chains, is presented in Figure~\ref{fig:ads-diagram}. We can see that, at least for the range of values sampled here, there is a rather linear relationship between these parameters independently of the magnetic or nonmagnetic nature of the filament. However, the adsorption of the MF is observed at slightly higher temperatures than its nonmagnetic counterpart. This can be attributed to the increased backbone stiffness led by the dipole-dipole interactions \cite{2013-sanchez-a}. Without regard the origin of the backbone rigidity, semiflexible polymer-like chains are known to adsorb on attractive surfaces at higher temperatures than their flexible counterparts \cite{1979-birshtein-bp, 2001-sintes, 2013-hsu-mm} due to their lower configurational entropy and, thus, to their lower average entropic repulsion with walls. Here, the anisotropic nature of the dipole-dipole interaction and the coupling between the dipole orientations and the chain backbone can lead only to a decrease of the chain configurational entropy with respect to the nonmagnetic case.

The adsorption transition of a filament involves the change from three- to two-dimensional structures. However, this is not the only structural change that this type of chain-like systems experience on cooling. First, the drecrease of thermal fluctuations tends to reduce the stretching and bending of the bonds, making the backbone locally smoother \cite{2011-sanchez-sm}. As pointed above, in MFs the latter is favored by the dipole-dipole interactions. Besides, MFs also may experience a magnetic flux closure transition on cooling, changing from open to ring-like structures \cite{2011-sanchez-sm, 2013-sanchez-a}. In order to determine how these three effects interact in our system, we computed two standard structural parameters: the radius of gyration
\begin{equation}
 R_g = \left [ \frac{1}{N_p} \sum_{i=1}^{N_p} (\vec r_i - \langle \vec r \rangle)^2\right ]^{1/2} ,
 \label{eq:Rg}
\end{equation}
where $\langle \vec r \rangle = \sum_{k=1}^{N_p} \vec r_k / N_p$, and the end-to-end distance
\begin{equation}
 R_{ee} = \left \| \vec r_1 - \vec r_{N_p}\right \| ,
\end{equation}
calculated from the positions of the filament particles, $\vec r_i$, being $\vec r_1$ and $\vec r_{N_p}$ those corresponding to the chain ends. Note that for a MF with dipole moments coupled to the chain backbone, $R_{ee}$ is basically proportional to its net magnetization \cite{2011-sanchez-sm, 2013-sanchez-a}. Figure~\ref{fig:rads-nofield} shows the dependence of the averages of these parameters on $T$ for both cases, $\mu^2=0$ and $\mu^2=5$, and for different surface attraction strengths. Such averages have been obtained also from WHAM calculations. At high temperatures both, MFs and nonmagnetic chains in a desorbed state, adopt a random coil structure. As the temperature is decreased the chains experience an important straightening, which is evidenced by the significant growth of $\langle R_g \rangle$ and $\langle R_{ee} \rangle$ and illustrated respectively by the snapshots (iv), (iii) and (ii) in Figure~\ref{fig:rads-nofield}. Nonmagnetic chains keep experiencing such straightnening as the temperature is further reduced to its minimum sampled value, even after the characteristic adsorption temperature, $T_{\mathrm{ads}}$, is reached (see snapshot (i) in the same Figure). Straightening of MFs, however, happens only for temperatures above a certain value. At temperatures below such limit, however, they show an abrupt drop of both, $\langle R_g \rangle$ and $\langle R_{ee} \rangle$, associated to a prominent peak in the fluctuations of these parameters (shown in the insets). Such drop signals the adoption of a closed loop structure driven by the dipole-dipole interactions, as illustrated by snapshots (v) and (vi). Taking the position of the fluctuation peaks as the characteristic temperature of such closure transition, $T_{\mathrm{dip}}$, we can see that its value is practically constant for all adsorbed configurations, $T_{\mathrm{dip}} \approx 0.6$. In the case of the weakest sampled adsorption strength, $\bar \epsilon_{\mathrm{s}} \approx 0.14$, for which the full adsorption was not reached within the sampled interval of temperatures, the desorbed MF also experiences a closure transition but at a lower temperature, $T_{\mathrm{dip}} \approx 0.5$. This can be explained by the higher configurational entropy of the desorbed filament compared to its adsorbed state.

\begin{figure}[!t]
 \centering
 \includegraphics[width=8.5cm]{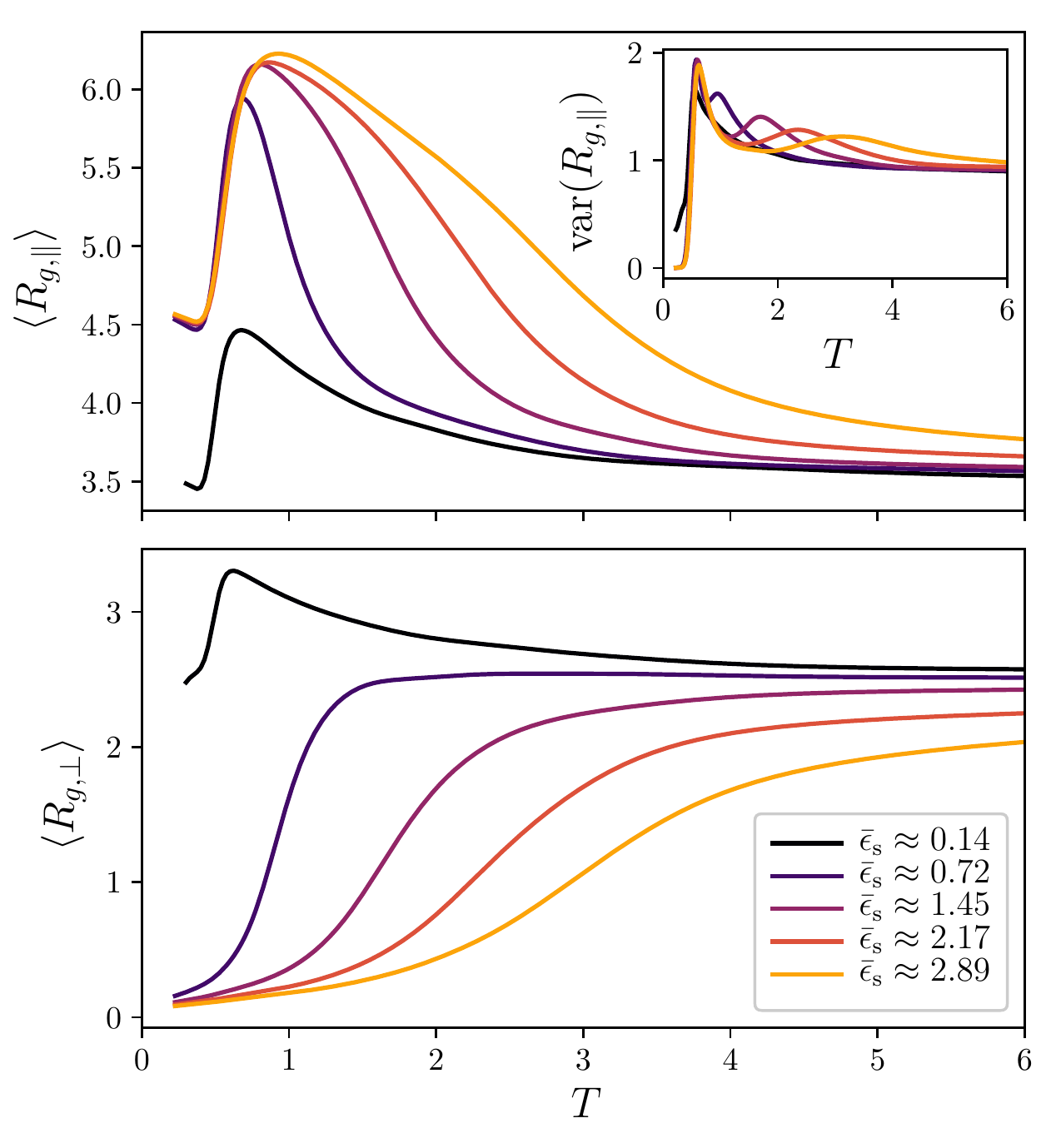}
 \caption{Average parallel (top panel) and perpendicular (bottom panel) components of the radius of gyration of the grafted MF ($\mu^2=5$) as a function of the temperature for different surface attraction strengths. Inset in top panel shows the fluctuations of the parallel component.}
 \label{fig:splitrads}
\end{figure}
The separation of the two transitions of the MF, adsorption and closure, found under not too weak surface attraction conditions, can be better visualized by splitting the components of the radius of gyration parallel and perpendicular to the surface, $R_{g,\parallel}$ and $R_{g,\perp}$. These parameters are obtained from Eq.~(\ref{eq:Rg}) by taking the corresponding components of the position vectors. Figure~\ref{fig:splitrads} shows their WHAM averages as a function of the temperature for each surface attraction strength. We can see that for $\bar \epsilon_{\mathrm{s}} \gtrsim 0.72$ the parallel component captures the initial chain straightening and its subsequent closure as $T$ is decreased, whereas the perpendicular component reflects the adsorption transition. The comparison of these curves evidences that both, straigthening and adsorption happen simultaneously for this range of parameters. Interestingly, the fluctuations of each $R_{g,\parallel}$ curve show two peaks, which signal the characteristic adsorption and closure temperatures. As expected, the trace of the adsorption transition is absent from the curves of system $\bar \epsilon_{\mathrm{s}} \approx 0.14$. On cooling, it also displays a much weaker straightening before the onset of its closure.

In summary, our results indicate that the closure transition of the MF is independent of its adsorption transition for values of $\bar \epsilon_{\mathrm{s}}$ large enough to impose $T_{\mathrm{ads}} > T_{\mathrm{dip}}$. Under such conditions the closure takes place with a two-dimensional constraining of the filament and we observe $T_{\mathrm{dip}} = \left ( T_{\mathrm{dip}} \right )_{\mathrm{adsorbed}} \approx$ const. For low values of $\bar \epsilon_{\mathrm{s}}$ one can expect the adsorption transition to take place for a MF already in its closed state, so that $T_{\mathrm{ads}} < T_{\mathrm{dip}} = \left ( T_{\mathrm{dip}} \right )_{\mathrm{desorbed}}$, with $\left ( T_{\mathrm{dip}} \right )_{\mathrm{desorbed}} < \left ( T_{\mathrm{dip}} \right )_{\mathrm{adsorbed}}$. Besides these qualitative considerations, the accurate characterization of structural transitions at very low temperatures may require more refined simulation approaches, being out of the scope of this work.

\subsubsection{Adsorption transition under tilted fields}
Once the structural behavior displayed by this system on cooling has been characterized in detail, we address the main point of this work: the control of the adsorption/desorption of the MF at constant temperature by means of static homogeneous external fields. In general, such fields will tend to align each individual dipole in their direction, leading to an overall straightening and orientation of the chain backbone, thus, reducing its configurational entropy and increasing its effective stiffness. Qualitatively, in one hand one can expect the external field to hinder the closure transition of the MF while, on the other hand, its presence may favor or even force the adsorption or desorption of the MF depending on its strength and orientation: a field with strong enough component pointing out of the plane of the attractive surface can force the desorption, whereas a strong field component pointing into or parallel to such plane may favor the adsorption. However, the decrease in the configural entropy of the filament induced by the field makes difficult to anticipate its quantitative effects on the adsorption/desorption transition. Thus, it is necessary to characterize such effects in order to understand how the field can be used to control such transition.

In the following discussion, we take $T=1$ as fixed reference temperature and we consider only surface attraction strengths that led to an adsorption within the sampled range of temperatures, \textit{i.e.}, $\bar \epsilon_{\mathrm{s}} \gtrsim 0.72$. All parameters presented below are calculated from direct sampling averages.

\begin{figure}[h]
 \centering
 \includegraphics[width=8.5cm]{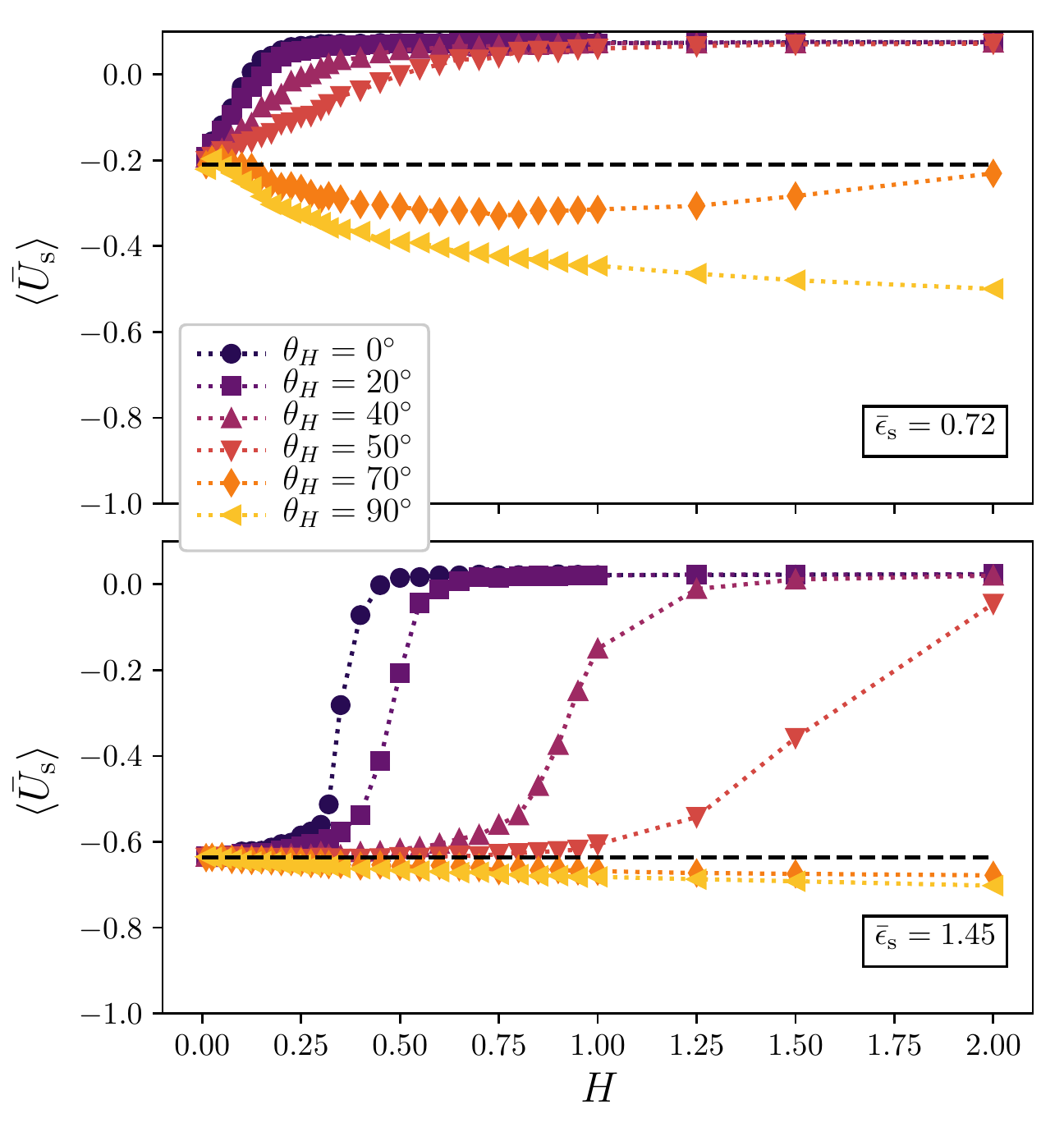}
 \caption{Average adsorption energy curves as a function of the field strength, $H$, for different field tilting angles, $\theta_H$, and two selected values of $\bar \epsilon_{\mathrm{s}}$. Dotted lines connecting the symbols and horizontal dashed lines are a guide to the eye. The latter correspond to the values of $\langle \bar U_{\mathrm{s}} \rangle$ at zero field.}
 \label{fig:ads-field}
\end{figure}
We start our analysis by examining the average surface energy as a function of field strength and orientation. Figure~\ref{fig:ads-field} shows such results for two selected values of $\bar \epsilon_{\mathrm{s}}$. From the position of the inflection points, we can see that, as expected, fields with small tilt angles easily force the complete desorption of the MF, requiring a weaker field to achieve it. The latter is signaled by $\langle \bar U_{\mathrm{s}} \rangle$ reaching its maximum saturation value, which in all cases is very close but not exactly equal to zero. The latter is a consequence of having the position of the end grafted particle permanently constrained within the range of interaction of the surface. Comparing the two absorption strengths, the saturation value of $\langle \bar U_{\mathrm{s}} \rangle$ is closer to zero for the stronger attraction, whereas the bigger deviation from zero at $\bar \epsilon_{\mathrm{s}} \approx 0.14$ simply reflects the stronger fluctuations of the position of the grafted end particle due to the smaller depth of the surface potential well. From the same comparison, one can also observe that the weaker is the adsorbed state at zero field, as signaled by a relatively high surface energy, the weaker is the field strength required to force the desorption. As $\theta_H$ increases, such forced desorption requires stronger fields up to a point in which the field inverts its effect and starts to favor the adsorption. Importantly, such characteristic angle decreases with increasing values of $\bar \epsilon_{\mathrm{s}}$.

The results shown in Figure~\ref{fig:ads-field} confirm the expected monotonous dependence of the field effects on its tilting angle. However, what happens for a fixed angle when the field strength is changed is a more subtle question, as the nonmonotonic profile of the curve corresponding to $\bar \epsilon_{\mathrm{s}} \approx 0.72$, $\theta_H = 70^{\circ}$ evidences: for weak strengths the adsorption is favored as $H$ grows, but only up to a certain point after which any further increase favors the desorption. Such dependence on the field strength can be explained by the interplay between the filament configurational entropy and the two main terms of the energy, \textit{i.e.}, the magnetic and the surface interaction terms. Since at $\theta_H = 70^{\circ}$ the main component of the field is parallel to the surface, magnetic interactions tend to not only decrease the overall configurational entropy but also to penalize the exploration of configurations occupying regions far from the surface. Note, however, that lateral configurational fluctuations with respect to the axis defined by the field direction can not be fully hindered at any field strength, which means that under large tilt angles the surface remains entropically reachable by the filament. However, any large tilt angle $0 \ll \theta_H < 90^{\circ}$ still puts a magnetic energy penalty on fully adsorbed states due to the misalignment of the field and the surface. At weak fields such penalty is relatively small and the interaction with the surface can dominate. The latter benefits from the initial decrease in the fluctuations led by the growth of the field strength, being able to overcompensate the increase of magnetic energy for the adsorbed state. This corresponds to the observed initial enhacement of the adsorption led by weak growing fields. At some field strength such energy balance saturates and finally inverts, becoming the interaction with the field the dominant one at high $H$. Importantly, even though for the set of parameters sampled here we observe this nonmonotonic behavior only in one case, the reasoning of its explanation can also be applied to other large values of $\theta_H$ and even to systems with different $\bar \epsilon_{\mathrm{s}}$. For instance, one can expect the curve for $\theta_h = 70^{\circ}$, $\bar \epsilon_{\mathrm{s}} \approx 1.45$ to also invert its trend at very large values of $H$.

\begin{figure}[h]
 \centering
 \subfigure{\label{fig:ads-fields}\includegraphics[width=8.5cm]{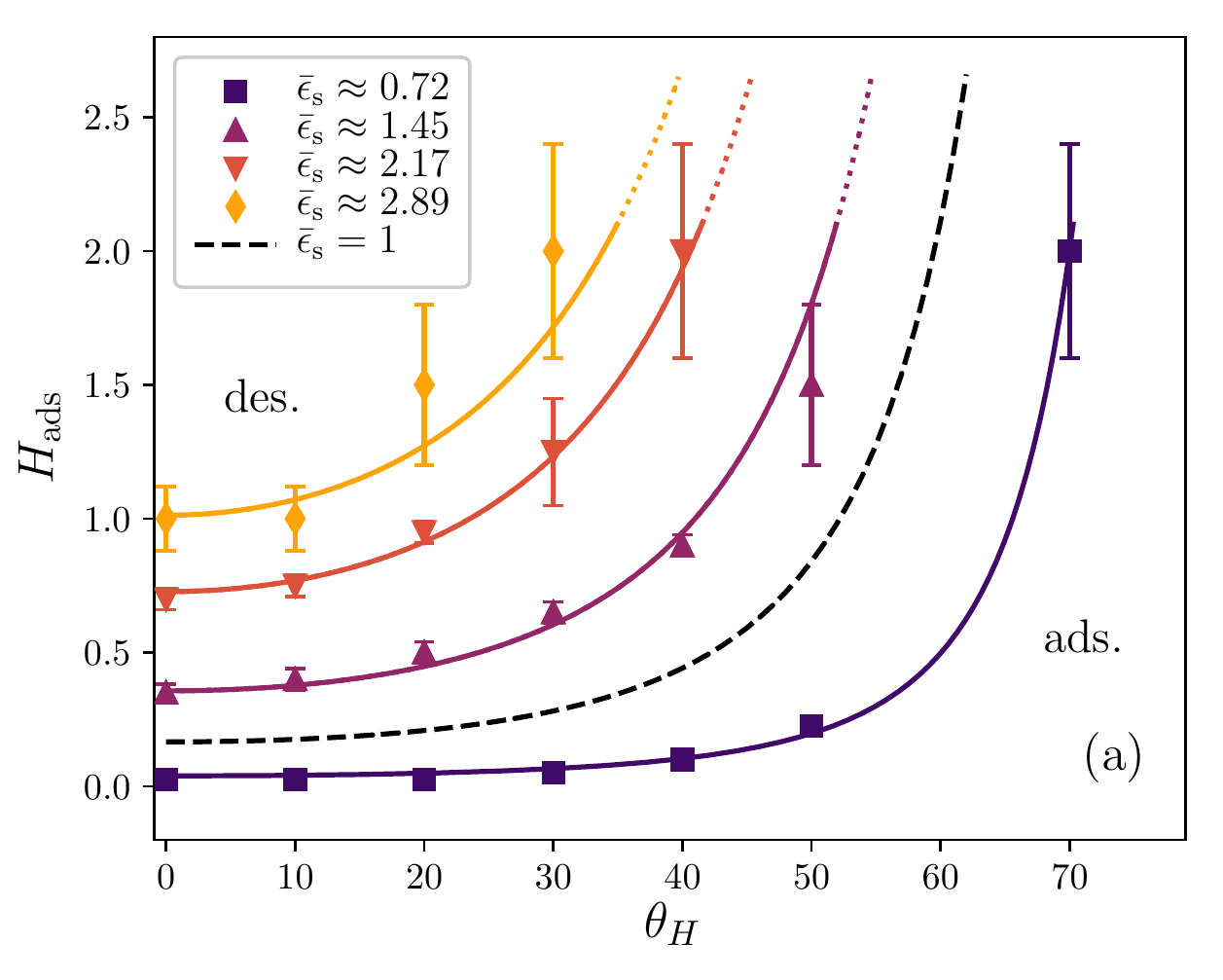}}
 \subfigure{\label{fig:ads-fields-master}\includegraphics[width=8.5cm]{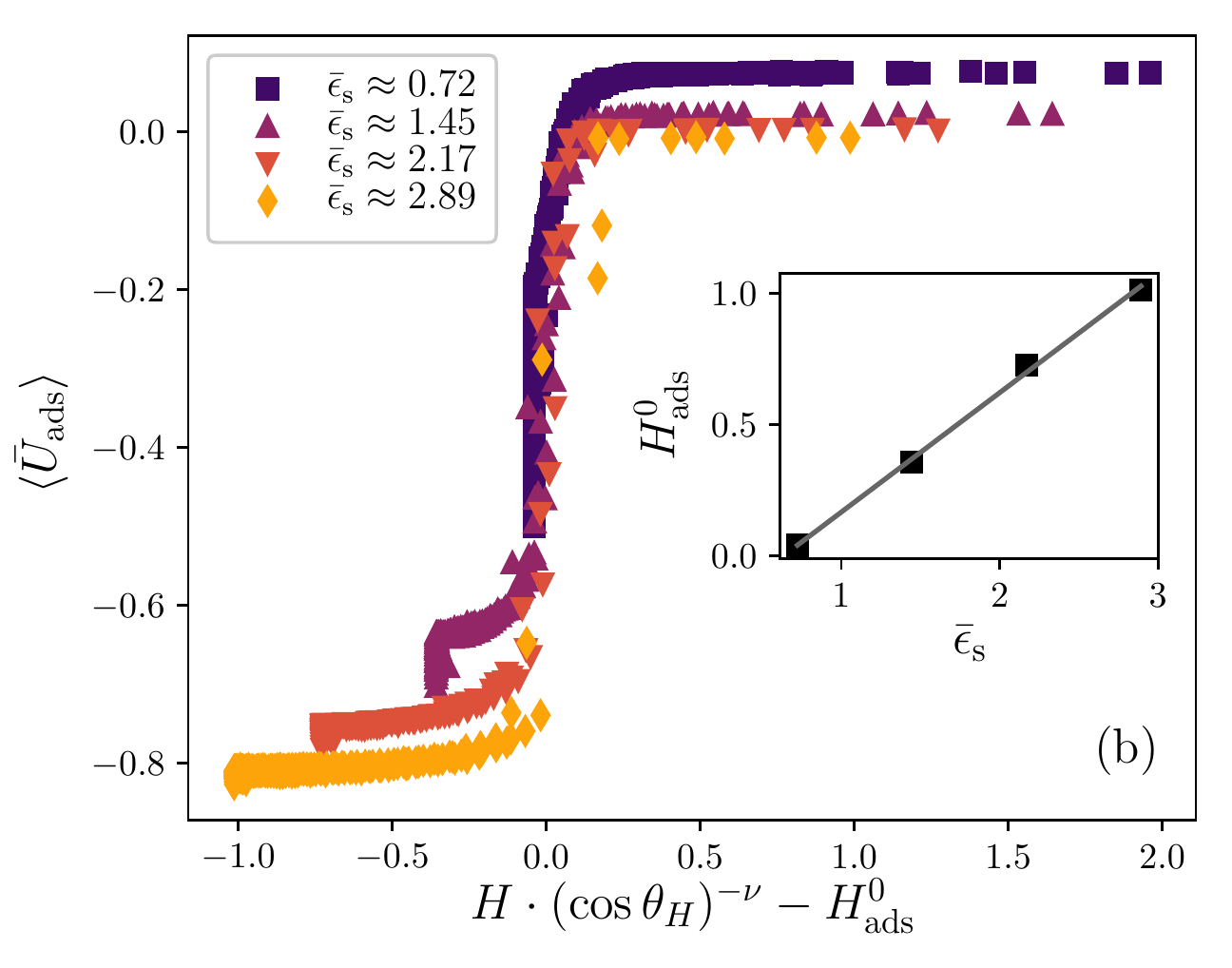}}
 \caption{(a) Diagram of characteristic boundaries between adsorbed and desorbed states obtained from the maxima of fluctuations of $\langle \bar U_{\mathrm{s}} \rangle$ (symbols) and corresponding least-squares fitting of Eq.~(\ref{eq:powerlaw}) (solid lines). Dotted segments are extrapolations beyond the range of field strengths sampled here, which are considered meaningful only for $\bar \epsilon_{\mathrm{s}} \gtrsim 1.45$. Dashed line is the predicted curve for $\bar \epsilon_{\mathrm{s}} =1$, obtained from the same expression by interpolation of the corresponding zero field prefactor, $H_{\mathrm{ads}}^0$. (b) Collapse of the field-induced adsorption-desorption curves for all sampled parameters. Inset shows the linear dependence of the fitted values of $H^0_{\mathrm{ads}}$ on $\bar \epsilon_{\mathrm{s}}$.}
\end{figure}

In order to analyze the switching of the field effects on the adsorption behavior, here we focus on the regions of monotonous response to weak fields and determine the boundaries between the adsorption and desorption regimes as a function of the tilt. Analogously to the analysis of the transitions in temperature discussed above, we take the position of the maximum of the fluctuations of $\langle \bar U_{\mathrm{s}} \rangle$ as the point that represents the characteristic boundary between the adsorbed and desorbed states. Symbols in Figure~\ref{fig:ads-fields} show the field strength at such characteristic boundary, $H_{\mathrm{ads}}$, obtained for each $\bar \epsilon_{\mathrm{s}}$ as a function of the tilting angle. We can observe that $H_{\mathrm{ads}}(\theta_H)$ exhibits a regular trend in all cases, with a slight growth for small tilting angles that becomes very steep for larger ones. This suggests that $H_{\mathrm{ads}}$ may follow a simple unique function of $(\theta_H)$. According to the discussion at the beginning of this Section, it is reasonable to assume that the relevant magnitude controlling the desorbing effect of the field is its component perpendicular to the adsorbing surface. Therefore, we assumed the following power law for $H_{\mathrm{ads}}$:
\begin{equation}
 H_{\mathrm{ads}} = H_{\mathrm{ads}}^0 (\cos \theta_H)^{-\nu},
 \label{eq:powerlaw}
\end{equation}
where $H_{\mathrm{ads}}^0$ is the characteristic field for $\theta_H=0$. We performed a least-squares fit of Eq.~(\ref{eq:powerlaw}) to all simulation datasets in Figure~\ref{fig:ads-fields}, obtaining a single fitted exponent $\nu=3.67 \pm 0.34$. The results of this fitting for each $\bar \epsilon_{\mathrm{s}}$ are shown in the same figure as solid lines. Note that curves corresponding to $\bar \epsilon_{\mathrm{s}} \gtrsim 1.45$ separate states of strong adsorption from desorbed configurations, whereas the curve for $\bar \epsilon_{\mathrm{s}} \approx 0.72$ bounds weakly adsorbed states only. In general, one can not expect a strong adsorption taking place for $\bar \epsilon_{\mathrm{s}} < T$. Such limiting condition is represented by the dashed curve in Figure~\ref{fig:ads-fields}, which shows Eq.~(\ref{eq:powerlaw}) with the value of $H_{\mathrm{ads}}^0$ corresponding to $\bar \epsilon_{\mathrm{s}} = T = 1$, as obtained by interpolation. Finally, it is worth mentioning that the extrapolation of these fitted curves can be considered valid as far as they do not reach relatively large tilts. This is the case of $\bar \epsilon_{\mathrm{s}} \gtrsim 1.45$ but not of $\bar \epsilon_{\mathrm{s}} \approx 0.72$. As discussed above, the latter already exhibits a nonmonotonic response at $\theta_H = 70^{\circ}$, thus for bigger tilts Eq.~(\ref{eq:powerlaw}) is not expected to hold for such a weak surface attraction strength.

The results presented above suggest that, under control of the external field, adsorption/desorption behavior can be described by a single master curve. This can be evidenced by performing a rescaling of all the transition curves obtained in our simulations according to fitted Eq.~(\ref{eq:powerlaw}). Figure~\ref{fig:ads-fields-master} shows the result of such rescaling for all available datasets, \textit{i.e.}, for all measured fields strengths, desorption-inducing tilting angles and the four considered surface attraction strengths. Apart from the expected $\bar \epsilon_{\mathrm{s}}$-dependent differences in the saturation values, there is a collapse of all data into a characteristic transition master curve that captures the effects of the external field on this system. Finally, as shown in the inset of Figure~\ref{fig:ads-fields-master}, $H^0_{\mathrm{ads}}$ values obtained from the fitting of Eq.~(\ref{eq:powerlaw}) show a linear dependence on $\bar \epsilon_{\mathrm{s}}$.

\section{Conclusions}
In this work we have studied, by means of computer simulations with a mesoscale model, the parameters that determine the equilibrium structural behavior of a flexible magnetic filament made of ferromagnetic nanoparticles and grafted to an attractive flat surface.

First, we characterized the different structural transitions that the filament experiences on cooling: backbone straightening, magnetic flux closure transition and adsorption on the attractive surface. We have shown that, for surface attraction strengths that compare to the thermal fluctuations at room temperature or above, the straightening and the adsorption transition take place simultaneously, whereas the closure transition requires lower temperatures. We also evidenced that the magnetic filament adsorbs at slightly higher temperatures than its nonmagnetic equivalent chain due to its effective increased rigidity induced by dipole-dipole interactions.

Finally, we studied the conditions to control the adsorption/desorption of the magnetic filament by means of static homogeneous magnetic fields. We have shown that the state of the filament can be effectively controlled by both, the strength and the orientation of the applied field. Filament desorption can be easily forced by fields perpendicular to the adsorbing surface. As the field tilting angle with respect to the normal increases, stronger fields are needed to force the desorption, until a maximum angle is reached and the effect of the field is inverted, forcing the adsorption. Importantly, the characteristic field strengths and tilting angles that separate the adsorbed and desorbed states are related by a simple power law whose prefactor depends linearly on the surface attraction strength. Therefore, the field-induced adsorption/desorption of the filament is fully represented by a transition master curve. This fundamental characterization may be essential for the future design of field-switchable micro- or nanofluidic devices based on magnetic filaments.

We consider the results presented here as a preliminary step on the way to the design of magnetically controlled filtering microfluidic devices based on MFs. Future works with this perspective will require to include hydrodynamic interactions in order to study the dynamic response and nonequilibrium properties of these systems.

\section*{Acknowledgements}
This research was supported by the Russian Science Foundation, Grant No.19-12-00209. T.S. acknowledges support by the Spanish AEI/MCI/FEDER(UE), Grant No. RTI2018-095441-B-C22 and The Maria de Maeztu R$\&$D Program (MDM-2017-0711). Simulations were carried out at the Vienna Scientific Cluster (VSC).

\end{document}